# Inverse problem solver for multiple light scattering using modified Born series

Moosung Lee[1,2†], Herve Hugonnet[1,2†], and YongKeun Park[1,2,3*]

[1] Department of Physics, Korea Advanced Institute of Science and Technology (KAIST), Daejeon 34141, South Korea;

[2] KAIST Institute for Health Science and Technology, KAIST, Daejeon 34141, South Korea;

[3] Tomocube Inc., Daejeon 34109, South Korea

[†] These authors equally contributed to the work.

*corresponding authors: Y.K.P (yk.park@kaist.ac.kr)

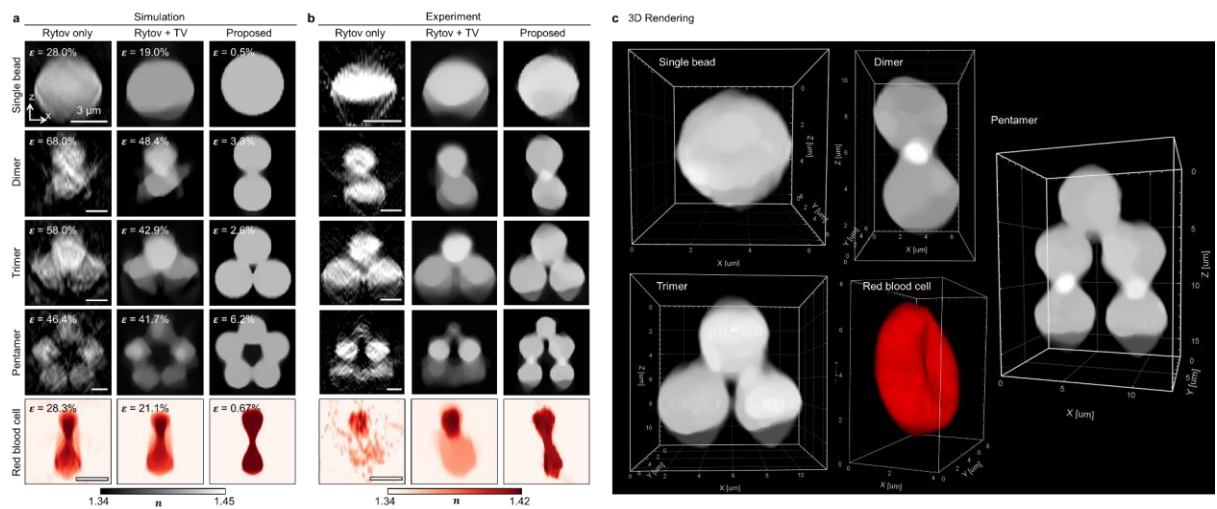

**ABSTRACT**

The inverse scattering problem, whose goal is to reconstruct an unknown scattering object from its scattered wave, is essential in fundamental wave physics and its wide applications in imaging sciences. However, it remains challenging to invert multiple scattering accurately and efficiently. Here, we exploit the modified Born series to demonstrate an inverse problem solver that efficiently and directly computes inverse multiple scattering without making any assumptions. The inversion process is based on a physically intuitive approach and can be easily extended to other exact forward solvers. We utilised the proposed method in optical diffraction tomography and numerically and experimentally demonstrated three-dimensional reconstruction of optically thick specimens with higher fidelity than those obtained using conventional methods based on the weak scattering approximation.

# INTRODUCTION

Scattering theory describes the interaction of wave fields with matter and is used in diverse fields of physics and engineering applications. Scattering theory can mainly be divided into two categories of problems: forward and inverse. The forward problem involves computing the scattered field from a known structured medium, whereas the inverse problem goal is to obtain the structured media from a known scattered field. Several forward solvers have been established and are widely used; the finite-difference time-domain (FDTD) method is one example[1]. In contrast, because the inverse problem is ill-posed and has complex computation requirements, it is considered challenging. Even with various approximations and assumptions, the inverse problem is more difficult to solve compared with the forward problem.

There are ongoing efforts to develop accurate and efficient inverse solvers and some of the solutions proposed have shown promise in various wave physics and imaging applications. Conventional inverse scattering methods impose weak scattering assumptions, such as the Born[2] or Rytov[3] approximation, and neglect high-order multiple scattering. Although they are computationally efficient, the inherent assumptions of the methods conflict with the optical properties of the samples, limiting their accuracy and applicability. To take into account the effect of multiple scattering, recent studies have exploited forward propagation methods, including the beam propagation method[4,5], angular spectrum method[6], the split-step non-paraxial method[7] and the scalar Lippmann–Schwinger model[8]. Although these methods provide more accurate results than weak scattering models, they still involve assumptions such as paraxiality or scalar diffraction. A general theoretical framework for accurate and fast inverse scattering models is required.

In this paper, we present an efficient and accurate inverse problem solver for the inverse scattering problem. Our proposed method exploits the modified Born series, which is currently one of the most efficient forward solvers for simulating wave propagation in inhomogeneous media[9]. Furthermore, we exploit the designed framework for optical diffraction tomography (ODT), in which accurate and efficient tomographic refractive index (RI) reconstruction of optically thick samples has been considered extremely challenging. In both simulations and experiments, our proposed method accurately captured the volumetric information of optically thick specimens, including clusters of microspheres, live cells, and human biological tissue.

## Results

**The modified Born series**

We exploited a modified Born series to solve the inverse scattering problem accurately and efficiently (Fig. 1). Originally, the modified Born series was set out to efficiently solve the forward problem of the Lippmann–Schwinger equation[9] (Fig. 1a):

$$\psi_{out}(\mathbf{r}) = \psi_{in}(\mathbf{r}) + \int G(\mathbf{r}-\mathbf{r}')V(\mathbf{r}')\psi_{out}(\mathbf{r}')d\mathbf{r}', \qquad (1)$$

where $\psi_{in}(\mathbf{r})$ and $\psi_{out}(\mathbf{r})$ are the incident and output monochromatic fields, respectively; $G(\mathbf{r}) = \exp[ikr]/4\pi r$; $k = 2\pi n_m/\lambda$; $n_m$ and $\lambda$ are the medium RI and the wavelength, respectively; $V(\mathbf{r}) = k^2[n^2(\mathbf{r})/n_m^2 - 1]$ is the 3D scattering potential; and $n(\mathbf{r})$ is the sample RI. In the discrete space, where the region of space is divided into $N$ voxels, Eq. (1) can be linearized as follows:

$$\boldsymbol{\Psi}_{out} = \boldsymbol{\Psi}_{in} + \mathbf{G}\mathrm{diag}(\mathbf{v})\boldsymbol{\Psi}_{out}, \qquad (2)$$

where $\boldsymbol{\Psi}_{out}$, $\boldsymbol{\Psi}_{in}$ and $\mathbf{v} \in \mathbb{R}^{N\times 1}$ are vectorised forms of $\psi_{out}(\mathbf{r})$, $\psi_{in}(\mathbf{r})$, and $V(\mathbf{r})$, respectively; and $\mathbf{G} = \mathbf{U}^\dagger \mathrm{diag}[(\mathbf{q}^2 - k^2)^{-1}]\mathbf{U} \in \mathbb{R}^{N\times N}$ represents the convolution operator using Green's function. $\mathbf{U}$ represents the discrete Fourier transform from the image space to the Fourier space with a spatial angular frequency vector, $\mathbf{q}$. The forward problem can then be expressed using matrix inversion:

$$\boldsymbol{\Psi}_{out} = [\mathbf{I} - \mathbf{G}\mathrm{diag}(\mathbf{v})]^{-1}\boldsymbol{\Psi}_{in}, \qquad (3)$$

where $\mathbf{I}$ denotes the identity matrix. Equivalently, the output scattered field $\boldsymbol{\Psi}_s = \mathbf{G}\mathrm{diag}(\mathbf{v})\boldsymbol{\Psi}_{out}$ can be computed as:

$$\Psi_{out} = [\mathbf{I} - \mathbf{G}\mathrm{diag}(\mathbf{v})]^{-1}\mathbf{Gs}, \qquad (4)$$

where $\mathbf{s} = \mathrm{diag}(\mathbf{v})\Psi_{in}$ is the source field. The measured field at the detector plane can be modelled as $\mathbf{y} = \mathbf{P}\Psi_{out}$, where $\mathbf{P} = \mathbf{U}^{\dagger}\mathrm{diag}[P(\mathbf{q})]\mathbf{U}$ represents a projection operator that limits the sample information at the pupil plane by the pupil function $P(\mathbf{q})$. When $N$ is large, direct matrix inversion is computationally inefficient. For efficient computation, the conventional Born series expands the equation in the Taylor series (Fig. 1b):

$$\Psi_s = \sum_{n=0}^{\infty}[\mathbf{G}\mathrm{diag}(\mathbf{v})]^n \mathbf{Gs}, \qquad (5)$$

which often diverges because $\mathbf{G}\mathrm{diag}(\mathbf{v})$ is generally outside the radius of convergence. Osnabrugge et al.[9] resolved this problem by modifying the equation into a convergent form:

$$\Psi_s = \sum_{n=0}^{\infty}[\mathbf{M}(\mathbf{v})]^n \mathbf{G}_\eta \mathbf{s}, \qquad (6)$$

where $\mathbf{M}(\mathbf{v}) = \mathbf{I} + i\mathrm{diag}(\mathbf{v} - i\eta)[\mathbf{G}_\eta\mathrm{diag}(\mathbf{v} - i\eta) - \mathbf{I}]/\eta$, $\eta$ is an arbitrary value larger than $\max(|\mathbf{v}|)$, and $\mathbf{G}_\eta = \mathbf{U}^{\dagger}\mathrm{diag}[(\mathbf{q}^2 - k^2 - i\eta)^{-1}]\mathbf{U}$. The modified Green's function can be physically interpreted as an evanescent spherical wave. Correspondingly, Eq. (6) can be understood as a modified Huygens' principle, where individual scattering points generate evanescent spherical waves. The exponential decay of the evanescent spherical wave is then compensated by introducing a gain, $-i\eta$, into the scattering potential.

We confirmed the accuracy of the modified Born series in a simulation by comparing it with the results obtained from the Mie theory[10,11] and FDTD method (Fig. 1c). To consider the vectoral properties of light, a dyadic formula can be used in the forward model (see *Methods*). In a $501 \times 501 \times 81$ gridded volume, a 3-μm-diameter bead ($n = 1.461$) in water ($n_m = 1.336$) was simulated to scatter a normally incident green plane wave ($\lambda = 532$ nm). The conventional Born series diverged, whereas our implemented algorithm and the FDTD method resulted in scattered fields in agreement with the results of Mie theory. Importantly, the computation time of the modified Born series was 36.6 s when we used a graphical processing unit (GPU; GeForce GTX1070 Ti), which is two orders of magnitude lower than that required by the FDTD method (2,414 s).

**Inverse scattering algorithm**

We aimed to find an unknown vectorised scattering potential, $\mathbf{v}'$, from a measured scattered field by minimising the cost function $\varepsilon(\mathbf{v}')$,

$$\mathbf{v}' = \arg\min_{\mathbf{v}' \in \mathbf{R}^{N \times 1}} \varepsilon(\mathbf{v}') = \arg\min_{\mathbf{v}' \in \mathbf{R}^{N \times 1}} \frac{1}{2}\|\Delta \mathbf{y}'(\mathbf{v}')\|_2^2, \tag{7}$$

where $\Delta \mathbf{y}'(\mathbf{v}') = \mathbf{y}'(\mathbf{v}') - \mathbf{y}$; $\mathbf{y}'(\mathbf{v}')$ and $\mathbf{y}$ are the expected and measured fields at the detector, respectively (Fig. 1d). The expected field at the detector can be expressed using Eq. 6 as

$$\mathbf{y}'(\mathbf{v}') = \mathbf{P}[\mathbf{I} - \mathbf{G}\mathrm{diag}(\mathbf{v}')]^{-1}\boldsymbol{\Psi}_{in}. \tag{8}$$

Here, pupil operator $\mathbf{P}$ is defined by Eq. (9) to represent the experimental measurement conditions in which the scattered field is low-pass filtered, refocused to the focal plane, and affected by aberrations due to the optical system, including those due to demagnification by the lens[12]

$$\mathbf{P} = \mathbf{U}^{\dagger}\mathrm{diag}\left[\mathrm{rect}\left(\frac{q_{xy}}{2\pi \cdot \mathrm{NA}/\lambda}\right)\exp[-ik_z z_{far}]A(\mathbf{q})\right]\mathbf{U}\mathrm{diag}\left[\delta(z - z_{far})\right], \tag{9}$$

where $A(\mathbf{q})$ represents the Fourier space aberrations; $q_{xy}$ and $k_z = (k^2 - q_{xy}^2)^{1/2}$ are the lateral and axial wavenumber respectively. We assume that the aberrations of the fields have been corrected before inversion; thus, $A(\mathbf{q})$ can be omitted.

Similar to the procedure used in the seminal study by Soubies et al.[13], an analytic form of the cost function gradient can be found (Matrix Cookbook[14] Eq. 247,136,232),

$$\nabla_{\mathbf{v}'}\varepsilon = \mathbf{D}_m(\mathbf{v}')^{\dagger}\Delta \mathbf{y}'(\mathbf{v}'), \tag{10}$$

where $\mathbf{D}_m(\mathbf{v}') = \nabla_{\mathbf{v}'}[\Delta \mathbf{y}'(\mathbf{v}')^T]$ is an $N \times N$ Jacobian matrix of the error function. Inserting Eq. (8) into Eq. (10), we obtain (Matrix Cookbook[14] in Eq. 59),

$$\nabla_{\mathbf{v}'}\varepsilon = [\mathbf{P}[\mathbf{I}-\mathbf{G}\text{diag}(\mathbf{v}')]^{-1}\mathbf{G}[\mathbf{I}-\mathbf{G}\text{diag}(\mathbf{v}')]^{-1}\text{diag}(\mathbf{\Psi}_{in})]^{\dagger}\Delta\mathbf{y}'(\mathbf{v}')$$
$$= \text{diag}(\mathbf{\Psi}'^{*})[[\mathbf{I}-\mathbf{G}\text{diag}(\mathbf{v}')]^{-1}\mathbf{G}]^{\dagger}\mathbf{P}^{\dagger}\Delta\mathbf{y}'(\mathbf{v}') \quad , \tag{11}$$

where $\mathbf{\Psi}'$ is the field in $\mathbf{v}'$ under illumination $\mathbf{\Psi}_{in}$. Note that $[\mathbf{I} - \mathbf{G}\text{diag}(\mathbf{v}')]^{-1}\mathbf{G} = \mathbf{G}[\mathbf{I} - \text{diag}(\mathbf{v}')\mathbf{G}]^{-1}$ because

$$[\mathbf{I}-\mathbf{G}\text{diag}(\mathbf{v}')]\{[\mathbf{I}-\mathbf{G}\text{diag}(\mathbf{v}')]^{-1}\mathbf{G} - \mathbf{G}[\mathbf{I}-\text{diag}(\mathbf{v}')\mathbf{G}]^{-1}\}[\mathbf{I}-\text{diag}(\mathbf{v}')\mathbf{G}] = \mathbf{0}. \tag{12}$$

Furthermore, because $\mathbf{G}^{\dagger} = \mathbf{G}$ and $\text{diag}(\mathbf{v}')^{\dagger} = \text{diag}(\mathbf{v}'^{*})$,

$$\left[[\mathbf{I}-\mathbf{G}\text{diag}(\mathbf{v}')]^{-1}\mathbf{G}\right]^{\dagger} = \left[\mathbf{I}-\mathbf{G}\text{diag}(\mathbf{v}'^{*})\right]^{-1}\mathbf{G}. \tag{13}$$

Using this result in Eq. (11), we obtain:

$$\nabla_{\mathbf{v}'}\varepsilon = \left[\text{diag}[\mathbf{\Psi}^{*}(\mathbf{v}')]\{\mathbf{I}-\mathbf{G}\text{diag}(\mathbf{v}'^{*})\}^{-1}\mathbf{GP}^{\dagger}\Delta\mathbf{y}'(\mathbf{v}')\right] = [\mathbf{\Psi}^{*}(\mathbf{v}') \odot \xi(\mathbf{v}')], \tag{14}$$

where $\odot$ denotes an elementwise product and $\xi(\mathbf{v}') = [\mathbf{I} - \mathbf{G}\text{diag}(\mathbf{v}'^{*})]^{-1}\mathbf{GP}^{\dagger}\Delta\mathbf{y}'(\mathbf{v}')$, which can be understood as the error-compensating field. For an aberration-corrected optical system, the error between the expected and measured field $\Delta\mathbf{y}'(\mathbf{v}')$ is first positioned by $\mathbf{P}^{\dagger}$ at the far-field plane where it was originally sampled and convoluted with the green function kernel $\mathbf{G}$ to form a 3D field. Then, this field is fed into the forward problem solver $[\mathbf{I} - \mathbf{G}\text{diag}(\mathbf{v}'^{*})]^{-1}$ (Eq. (6)). This means that the error between the measured and expected fields is back-propagated, similar to a phase conjugation experiment. Additionally, the medium absorption is changed into gain by the complex conjugation of the scattering potential, which compensates for the loss during forward propagation. Another interesting interpretation is that when starting from a homogeneous medium, the first iteration of the inversion is equivalent to using the Born approximation.

Our formulation here recapitulates the work of Soubies *et al.*[13] and provides a physically intuitive explanation of the inversion process. Furthermore, it enables the direct implementation of the inverse solver, given any forward solver. Here, we used the modified Born series algorithm presented in the previous section as a forward solver. The inversion algorithm can then be summarised as follows (Table 1):

| Inversion algorithm |
|---|
| **y** = measured fields; **v'** = Rytov RI estimate; $\Psi_{in}$ = illumination fields |
| **while** not converged |
|     [**y'**, $\Psi'_{3D}$] = simulate($\Psi_{in}$, **v'**) |
|     [~, $\xi_{err,\ 3D}$] = simulate(**y'** − **y**, flip(**v'**\*,3)) |
|     gradient = $i/(8\pi^2)$ $\Psi'_{3D}{}^*$ $\square$ flip($\xi_{err\ 3D}$, 3) |
|     **v'** = optimize(**v'**, gradient) |

Table 1. Pseudocode.

where [**y'**, $\Psi'_{3D}$] = simulate($\Psi_{in}$, **v'**) is the forward scattering solver returning both $\Psi'_{3D}$ (the 3D field in the simulation volume) and y' (the field at the detector plane). **v'** = optimise(**v'**, gradient) is the gradient-based optimisation step. Finally, flip(**v'**,3) flips the volume upside down to enable back propagation. More details about the inverse solver are provided in *Methods*.

We created an initial estimate of the scattering potential using the first-order Rytov approximation[15]. To enforce prior knowledge of the reconstructed scattering potential, we regularised the optimization problem with a total variation minimization algorithm[15] during the optimization step (Figs. 1e, f). Using the iterative proximal gradient descent method, we successfully reconstructed the 3D RI tomogram of the microspheres, red blood cells (Fig. 2), a chimeric antigen receptor T cell (Fig. 3), and tissue samples (Fig. 4). Further details about the code implementation can be found in *Methods*. The implemented MATLAB 2021a code is also available on GitHub (https://github.com/BMOLKAIST/Inverse_Solver).

**Tomographic reconstruction of microparticles in simulation and experiments**

We numerically and experimentally evaluated the performance of the proposed method using ODT to determine the sample RI distribution, $n(\mathbf{r})$. ODT is a 3D quantitative phase imaging technique, which reconstructed the 3D RI distribution of a sample from the measurements of multiple 2D optical fields using an inverse scattering algorithm[2,16,17]. We first assessed the reconstruction accuracy of the proposed method in simulations of various spherical bead multimers and red blood cells (Fig. 2a).

We compared the reconstruction performance of the proposed method with that of the first-order Rytov approximation[18]. The Rytov approximation requires accurate phase unwrapping because it assumes a spatially slowly varying phase of an output field. However, abrupt phase changes in the output field due to axially aligned samples impede robust phase unwrapping. As a result, the reconstructed RI tomograms using the Rytov approximation exhibited distorted artefacts even after total variation regularisation. In contrast, our method successfully recovered the scattering potential with high fidelity. Quantitatively, the relative mean-squared error (MSE) of the RI tomograms was an order of magnitude lower when our approach was used, thus validating its fidelity.

To further validate the proposed concept, we experimentally imaged and reconstructed the RI tomograms of the samples corresponding to the simulation conditions (see *Methods*). We specifically imaged colloidal suspensions of 5-μm-diameter silica microspheres in water and live mouse red blood cells (Fig. 2b). To increase the optical path delays, we axially aligned them using holographic optical tweezers[19-21]. As a result, the reconstructed RI tomograms using the Rytov approximation suffered from artefacts, even after regularisation. However, our method exhibited RI distributions that were closer to the expected 3D characteristics (Fig. 2c). Collectively, both simulations and experiments confirmed the feasibility of the proposed method.

**Tomography of 3D cell–cell interaction dynamics**

To validate the repeatable reconstruction performance of the proposed method, we assessed the 3D dynamics of cell–cell interactions. Specifically, we reconstructed the dynamics of a chimeric antigen receptor T cell, a genetically engineered immune cell for immunotherapy, attacking a target antigen-presenting cell, K562-CD19[22] (see *Methods*). The proposed method generated an 18-min 3D video sampled with the time intervals of 8 s (Fig. 3, see Supplementary Video 1). Compared with the conventional methods using the Rytov approximation (Figs 3a, b), our method visualised 3D RI tomograms with higher contrast of subcellular organelles and clearer 3D morphology (Figs 3c, d). This enhancement is particularly notable at the boundary between the two cells, where the microvilli of the T cell and synaptic cleft were more clearly resolved. Importantly, we observed consistently better performance of our proposed method over the 3D video data, which suggests the potential of our framework for accurate 3D quantitative analyses in cell biology[16].

**Tomography of optically thick biological samples**

We leveraged the reconstruction performance of the proposed method by reconstructing 3D RI tomograms of optically inhomogeneous biological samples with a thickness exceeding tens of micrometres. We reconstructed 3D RI tomograms of microalgae, *Chlorococcum oleofaciens* and *Pyropia yezoensis*, which exhibited strong RI inhomogeneities owing to the presence of thick cell walls, chloroplasts, and lipid droplets (Figs. 4a, b; see *Methods*). A comparison of reconstruction results showed that our proposed method clarified those characteristic structures of both cells in 3D with improved contrast. Interestingly, the cross-sectional images revealed that our method could resolve the vesicles exhibiting RI values larger than 1.50 and the top part of the cells, whereas the conventional methods could not. These overall inspections imply the importance of considering multiple light scattering in tomography of highly scattering cells.

Finally, we reconstructed 3D RI tomograms of a 200 μm-thick human pancreas tissue, whose reconstruction is highly challenging using the conventional ODT. Reconstruction of thick specimens requires large memory and a good initial estimate. We addressed the former by tiled reconstruction; in other words, we divided each raw data point into overlapping tiled subsets, reconstructed each subset, and reassembled it into a final blended tomogram. The latter was resolved using holographic refocusing of the obtained light field images to virtual focal planes using a diffraction kernel with 5-μm intervals[23]. As a result, we successfully reconstructed a 200 μm-thick human pancreas tissue over a lateral field-of-view of 107 × 107 μm$^2$ (Fig. 4c). Compared with the conventional methods based on the Rytov approximation, the proposed method successfully reconstructed the detailed 3D structures inside thick samples larger than 35 μm in thickness with significantly enhanced contrast. Taken together, our inverse problem solver is accurate, efficient, repeatable, and scalable for general applications where multiple scattering needs to be considered.

## Discussion

A dilemma of the inverse scattering problem has been the trade-off between computational accuracy and efficiency. In this study, we propose a rigorous but intuitive inversion theory and take advantage of the modified Born series to suggest that both accuracy and efficiency can be achieved simultaneously. Compared with previous approaches, our method efficiently computes the gradient of field errors in the presence of multiple scattering without strict assumptions. Our numerical and experimental results clearly support the generality of the proposed method for improving the reconstruction performance in optical diffraction tomography. It can also be readily applicable to existing ODT data to improve the quality of images as well as the precision and accuracy of RI values, allowing the systematic investigation of subcellular structures of unlabelled live biological cells[24] and the quantitative analyses of their dry-mass concentrations[25-27]. Also, the present method can also be extended to be implemented to other 3D quantitative phase imaging techniques. With these unique advantages, we anticipate the deployment of our framework in biomedical studies of thick specimens, such as the metabolism of *Caenorhabditis elegans*[28], 3D cell culture, tumor spheroid, early embryogenesis[29], and 3D histopathology[30].

An important next step would be to further optimise the computational speed. Exact forward solvers with linear complexity to the sample volume are actively researched[31] but are still impeded by their high complexity. Another available option is to impose additional prior knowledge about samples, such as RI homogeneity and finite RI range[32]. Indeed, we confirmed that simply imposing a finite RI range could accelerate and improve the reconstruction performance (see Supplementary Fig. 1).

We expect that our framework can be deployed not only in three-dimensional (3D) biomedical imaging, but can also be deployed in various application fields of scattering problems, including ptychography and inverse photonic designs, where backward modelling of multiple wave scattering is essential. In soft matter physics, for example, 3D reconstructions of birefringent materials can be achieved by considering polarisation and vectorised diffraction tomography[33,34]. In addition, our framework could be a potential alternative to the multi-slice approach in X-ray or electron ptychography for investigating volumetric ultrastructure[35,36]. Finally, along with the recent advances in nanophotonics, the proposed method may provide an efficient pipeline for the optimal design of silicon photonic devices[37], metamaterials[38], and metasurfaces[39,40]. In conclusion, we envisage that the proposed method will provide solutions to challenging scattering problems in various scientific and engineering research areas.

## Acknowledgements

This work was supported by KAIST UP program, BK21+ program, Tomocube, National Research Foundation of Korea (2015R1A3A2066550), and Institute of Information & communications Technology Planning & Evaluation (IITP) grant funded by the Korea government (MSIT) (2021-0-00745).

**Data availability** Our MATLAB 2021a implementation of forward and inverse models is available on GitHub (https://github.com/BMOLKAIST/Inverse_Solver).

## Conflict of interest

All authors have financial interests in Tomocube Inc., a company that commercialises ODT and quantitative phase imaging instruments and is one of the sponsors of this work.

## Author contributions

M.L. and H.H. implemented the algorithms, performed, and analysed the experiments. Y.P. supervised the project. All the authors were involved in writing the manuscript.

## Methods

**Dyadic formula of convergent modified Born series.** In an isotropic, linear, and dielectrically inhomogeneous medium, the inhomogeneous electromagnetic wave equation describes the vectoral scattering of light:

$$\left[\left(\nabla^2 + k^2\right)\mathbf{I} - \nabla\nabla^{\mathbf{T}}\right]\mathbf{E}(\mathbf{r}) = -V(\mathbf{r})\mathbf{E}(\mathbf{r}), \qquad (15)$$

where $\mathbf{E}(\mathbf{r})$ is the vector electric field. The solution of the equation is also described by the Lippmann–Schwinger equation (Eq. 1) but with the dipolar Green's function

$$\vec{\mathbf{G}}(\mathbf{r}) = \left[\mathbf{I} + \frac{1}{k^2}\nabla\nabla^{\mathbf{T}}\right]G(\mathbf{r}), \qquad (16)$$

which can be described as $\mathbf{G}_{dyad} = \mathbf{U}^\dagger \text{diag}[(\mathbf{q}^2 - k^2)^{-1}(\mathbf{I} - \mathbf{q}\mathbf{q}^T/k^2)]\mathbf{U}$.

**Forward solver.** We implemented a modified Born series based on acyclic convolution with a thin absorbing layer[41]. This type of implementation significantly suppresses wrap-around artefacts without requiring large zero padding. We chose the boundary thicknesses as $(6, 6, 6) \times \lambda/n_m$ or smaller along the $(x, y, z)$ directions. At each iteration, the field in the boundary region was multiplied by a linear ramp mask going from one in the sample to zero at the edge. To implement the acyclic convolution, we introduce a modified Fourier transform operator, $\mathbf{U}_\pm$, whose continuous version corresponds to

$$\mathbf{U}_\pm f(\mathbf{r}) = \int f(\mathbf{r}) e^{-i(\mathbf{q}\pm\delta\mathbf{k})\cdot\mathbf{r}} d\mathbf{r} = f(\mathbf{q} \pm \delta\mathbf{k}), \qquad (17)$$

which indicates the Fourier transform of $f(\mathbf{r})$ multiplied by a linear phase ramp. For a simulation volume $[L_x, L_y, L_z]$, $\delta\mathbf{k}$ was defined as $[1/L_x, 1/L_y, 1/L_z] \times \pi/2$. In the algorithm, any arbitrary convolution operator $\mathbf{F} = \mathbf{U}^\dagger \text{diag}[f(\mathbf{q})]\mathbf{U}$ was modified into $\mathbf{F}_\pm = (\mathbf{U}_\pm)^\dagger \text{diag}[f(\mathbf{q} \pm \delta\mathbf{k})]\mathbf{U}_\pm$. For example, the modified Born series operator, $\mathbf{M}$, is $\mathbf{M}_\pm = \mathbf{I} - \mathbf{\Gamma} + \mathbf{\Gamma}\mathbf{G}_{\eta\pm}\text{diag}(\mathbf{v} - i\eta)$, where $\mathbf{\Gamma} = i\text{diag}(\mathbf{v} - i\eta)/\eta$. The scattered electric field was then computed by recursion as $\mathbf{E} = \Sigma_{j=1}\mathbf{E}^{(j)}$, where $\mathbf{E}^{(1)} = \mathbf{\Gamma}\mathbf{G}_+\mathbf{s}$, $\mathbf{E}^{(2)} = \mathbf{M}_-\mathbf{E}^{(1)} + \mathbf{\Gamma}\mathbf{G}_-\mathbf{s}$, $\mathbf{E}^{(2j+1)} = \mathbf{M}_+\mathbf{E}^{(2j)}$, and $\mathbf{E}^{(2j+2)} = \mathbf{M}_-\mathbf{E}^{(2j+1)}$. The wrap-around errors at the edge of the computational volume changed the signs iteratively, thus effectively removing themselves during the iterations. We set the number of iterations such that the scattered evanescent waves completed pseudo-propagation to the ends of the computation volume.

**Inverse solver.** Here, we explain the origin of $i/8\pi^2$ in the pseudocode (Table 1). In the inversion algorithm, the error field is convoluted with the green function to obtain a 3D field for the backscattering simulation. The 2D Weyl expression of the dyadic Green function is[42]

$$\vec{\mathbf{G}}(\mathbf{r}) = \frac{i}{8\pi^2} \int_{-\infty}^{\infty}\int_{-\infty}^{\infty} \frac{1}{k_z}\left(I - \frac{\mathbf{k}_\pm \mathbf{k}_\pm^\dagger}{k^2}\right) e^{i\mathbf{k}_\pm \cdot \mathbf{r}} d^2\mathbf{q}_\perp, \qquad (18)$$

where $\mathbf{k}_\pm = (q_x, q_y, \pm k_z)$ with ± denoting the sign of $z$ and $k_z = (k^2 - q_x^2 - q_y^2)^{1/2}$. Here, $\exp[i\mathbf{k}_\pm \cdot \mathbf{r}]$ indicates a refocusing kernel and the remaining matrix operator, $[\mathbf{I} - \mathbf{k}_\pm\mathbf{k}_\pm^\dagger/k^2]/k_z$, is a transverse projection operator that ensures that the polarisation of the propagating light is perpendicular to the propagation direction and can be omitted for far field measurement, because those are composed of plane waves whose polarisation is already perpendicular to the propagation direction. This is followed by a convolution with $1/k_z$. Because $1/k_z$ is constant, and independent of the estimated RI, this term can be omitted by modifying the optimised cost function:

$$\varepsilon'(\mathbf{v}') = \frac{1}{2}\left\|\mathbf{U}^\dagger diag\left(\sqrt{k_{z1}}\right)\mathbf{U}\Delta\mathbf{y}'(\mathbf{v}')\right\|_2^2. \qquad (19)$$

Note that the modified cost function $\varepsilon'(\mathbf{v}')$ also reaches zero when the reconstructed RI equals the actual RI. Now, the gradient of the cost function is simply $i/(8\pi^2)\,\Psi'_{3D}{}^* \odot \mathrm{flip}(\Psi_{err\,3D}, 3)$.

Based upon the gradient computation, the proximal gradient method was used with multiple fields, one from the normal and the others from randomly chosen oblique angles. The number of used fields depended on the number of GPUs used; we chose 40 used fields when four GPUs were used, and 42 when seven GPUs were used. The T cell dataset was reconstructed by the stochastic gradient method with eight randomly chosen angles out of 49 holograms. We used the scalar Green's function for forward computation, which accelerated the computation while maintaining accuracy throughout the experiments. To denoise and deblur the reconstructed tomograms with accelerated speed, we implemented a total variation regularisation with fast iterative shrinkage-thresholding algorithm[15] (TV-FISTA) and optional non-negativity constraint. The voxel pitches were isotropically set to $\lambda/\alpha \text{NA}_{\text{obj}}$, where $\alpha = 3.6$ for the T cell dynamics, 3.7 for the tissue and 4 for the other datasets. The other parameters, including the internal and external iteration numbers, step sizes, and TV parameters, are summarized in Table 2.

**Far-field sampling conditions.** For the inverse solver to successfully converge, the boundary methods and operators used should be based on the same conditions at every step of the forward solver. When the cyclic boundary condition is used, the field can be simply sampled at the farthest simulation plane and then refocused by convolution with a cyclic digital refocusing kernel. Additionally, cyclic convolution operators were used, and the absorbing boundary layers in the modified Born series were set to zero pixels. The far-field propagation operator has the same form as in Eq. (9), with the range of integration limited by the finite field numerical aperture of the far field, for example, $(q_x^2 + q_y^2)^{1/2} < 2\pi \cdot \text{NA}/\lambda$.

The far-field projection operator became slightly more complicated when the acyclic boundary condition was used. Owing to the open boundary condition, sampling the field far from the sample would reduce the effective numerical aperture. For this reason, one iteration of the Born series (Eq. 2) was carried out with the green function replaced by the forward scattering far-field green function in order to obtain the far field at every position in the simulation volume. The far-field forward Green function is expressed as

$$\mathbf{G}_{\text{farfield}+} = \frac{i}{8\pi^2} \iint_{\sqrt{q_x^2+q_y^2} < 2\pi\frac{\text{NA}}{\lambda}} \frac{1}{k_z} \left( \mathbf{I} - \frac{\mathbf{k}_+ \mathbf{k}_+^\dagger}{k^2} \right) e^{i\mathbf{k}_+ \cdot \mathbf{r}} dq_x dq_y \quad (20)$$

and was evaluated using fine oversampling and making use of radial symmetry with the Hankel transform[43] to reduce computation cost. The acyclic convolution was then performed using zero padding. The field was then sampled directly at the focus plane. The same procedure was used to convert the far field into a 3D source field.

When boundary conditions were properly implemented, both the acyclic and cyclic models were robust and converged in all of our experiments. For a very highly scattering object, however, the optimisation could lead to a local minimum. The reconstructed RI map produced an almost identical transmitted field, although it was different from the original RI map. Therefore, the use of an initial RI estimate with the Rytov approximation is important.

**Experimental setup.** We collected experimental data from several ODT systems available in our laboratory[19,22,44], including a commercial ODT (HT-2H, Tomocube Inc., Republic of Korea). The common feature of these systems is single-shot, off-axis Mach–Zehnder holographic microscopy, which is equipped with a digital micromirror device (DMD) that controls the illumination angles of incident plane waves of continuous-wave lasers[45,46]. The relevant specifications of the setups include the illumination wavelength ($\lambda$), numerical aperture of an objective lens ($NA_{obj}$), scanning angles ($NA_{scan}$), and the number of illumination angles used ($N_{illum}$) (see Table 2). The details on the ODT setups, the field retrieval algorithm, and the conventional reconstruction algorithm can be found elsewhere[47,48]. Note that the setup used in Fig. 2 also included holographic optical tweezers to rotate samples and axially align the samples, as described elsewhere[19].

**Sample preparations.** The 5-μm-diameter silica bead multimers were purchased from Sigma-Aldrich, USA (44054-5ML-F). Live mouse red blood cells were kindly provided by Jieun Moon in KAIST. Blood extracted from a wild-type mouse through a syringe was immediately diluted in PBS to form an RBC solution. To prevent the samples from adhering to the imaging dish, the dish surface was coated with a 4% bovine serum albumin solution. Chimeric antigen T cells and CD19-positive K562 cells (K562-CD19; target cells) were kindly provided by Dr. Young-Ho Lee (Curocell Inc.). Both cells were prepared in complete RPMI medium with a number density of $2 \times 10^6$ cells/mL and mixed at a ratio of 1:1. *Chlorococcum oleofaciens* and *Pyropia yezoensis* suspended in Dulbecco's phosphate-buffered saline medium were kindly provided by Freshwater biological resource bank in Nakdonggang National Institute of Biological Resources and Laboratory for Biomimetic and Environmental Materials in POSTECH, respectively. The samples were mounted on a customised Petri dish for ODTs (Tomodish, Tomocube Inc.). The upper side of the dish was then covered with a No. 1.5H coverslip (20 × 20 mm$^2$, Marienfeld).

Human pancreatic tissue was prepared as a formalin-fixed paraffin-embedded tissue block. The block was sliced to a thickness of 200 μm, deparaffinized using three consecutive 10 min xylene baths, and finally mounted between two number zero coverslips using a mounting medium (Permount, RI = 1.53, Fisher Chemical).

**Figure legends**

**Figure 1 | Schematic of the proposed method. a,** The forward model finds the scattered field ($\Psi_{out}$) and the transmitted field (**y**) at the detector after passing through a pupil mask (**P**), given the incident field ($\Psi_{in}$) and a 3D scattering potential (**v**). **b,** Comparison between the conventional and modified Born series. The conventional Born series typically diverges, but the modified one always converges. **c**, Comparison of the estimated 3D $\psi_{out}$ with the Mie theory and FDTD results. **d,** The backward model iteratively finds unknown **v** given **y** and $\Psi_{in}$. This is computed by backpropagating errors using the modified Born series as a propagator. **e,** Comparison of reconstruction performance between the Rytov approximation and the proposed method. The tomogram reconstructed using the proposed method is rendered in 3D. **f,** Relative error curve per iteration. The final error at 200 iterations was 1.7%.

**Figure 2 | Reconstruction results of axially aligned freestanding specimens. a, b,** The reconstruction results in simulation (**a**). The simulated microspheres had the same diameter of 5 μm and RI of 1.42. The RI of the RBC phantom was 1.42. (**b**) Experimental results of axially aligned silica multimers and live mouse RBCs in water. **c,** Experimental data reconstructed by our proposed method are rendered in 3D.

**Figure 3 | Reconstruction results of dynamic 3D cell–cell interactions between chimeric antigen receptor T and K562-CD19 cells. a–c,** Cross-sections of tomograms reconstructed by (**a**) the Rytov approximation. (**b**) The Rytov approximation with TV regularization, and (**c**) our proposed method. Inset images: magnified images of white dashed squares. **d**, 3D rendered images of the results in **c**. The white arrows indicate either microvilli or a synaptic cleft.

**Figure 4 | Reconstruction results of optically thick specimens a–c,** Reconstruction results of (**a**) *Chlorococcum oleofaciens*, (**b**) *Pyropia yezoensis*, and (**c**) a 200-μm-thick human pancreas tissue. Each column represents cross-sectional images reconstructed by the Rytov approximation (first column), Rytov approximation with TV regularization (second column), and our proposed method (third

column). The last column represents 3D rendered images of the tomograms reconstructed by the proposed method.

**Table 1 | Pseudocode.**

**Table 2 | Setup specifications and used parameters.**

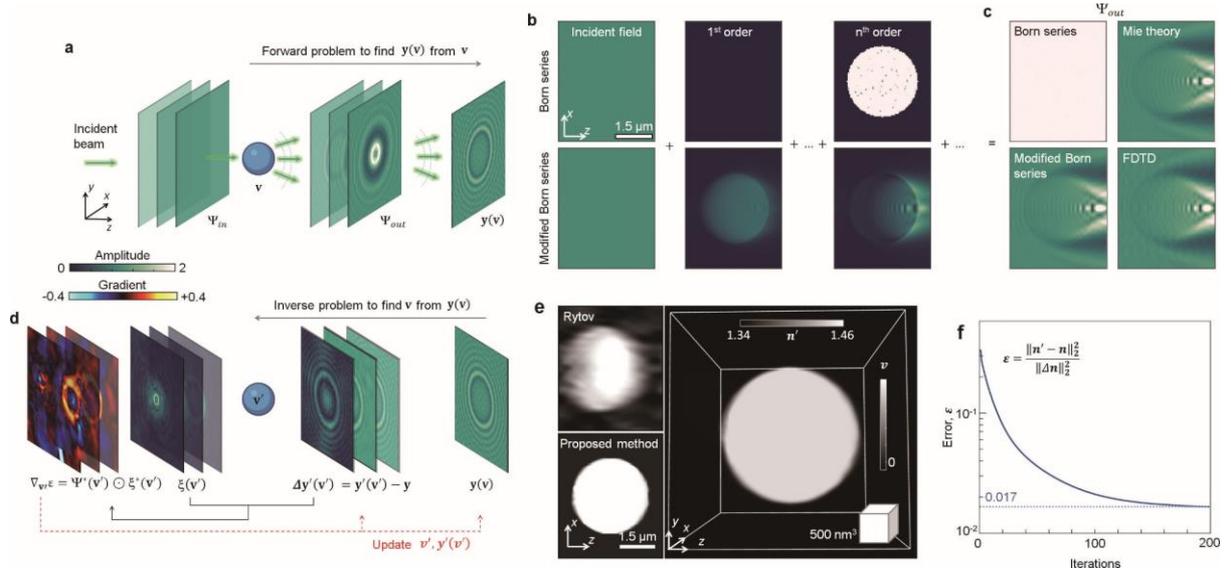

**Figure 1 | Schematic of the proposed method.**

**a,** The forward model finds the scattered field ($\Psi_{out}$) and the transmitted field (**y**) at the detector after passing through a pupil mask (**P**), given the incident field ($\Psi_{in}$) and a 3D scattering potential (**v**). **b,** Comparison between the conventional and modified Born series. The conventional Born series typically diverges, but the modified series always converges. **c,** Comparison of the estimated 3D $\Psi_{out}$ with the Mie theory and FDTD results. **d,** The backward model iteratively finds an unknown **v** given **y** and $\Psi_{in}$. This is computed by backpropagating errors using the modified Born series as a propagator. **e,** Comparison of reconstruction performance between the Rytov approximation and the proposed method. The tomogram reconstructed using the proposed method is rendered in 3D. **f,** Relative error curve per iteration. The final error after 200 iterations was 1.7%.

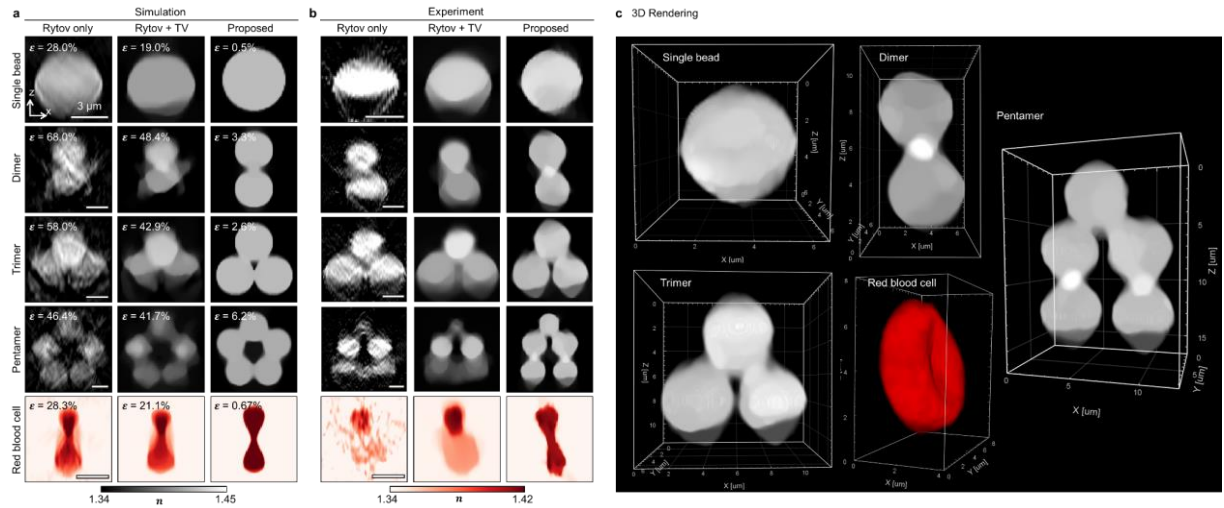

**Figure 2 | Reconstruction results of axially aligned freestanding specimens.**

**a, b,** Reconstruction results in simulation (**a**). The simulated microspheres had the same diameter of 5 μm and RI of 1.42. The RI of the RBC phantom was 1.42. (**b**) Experimental results of axially aligned silica multimers and live mouse RBCs in water**. c,** Experimental data reconstructed by our proposed method are rendered in 3D.

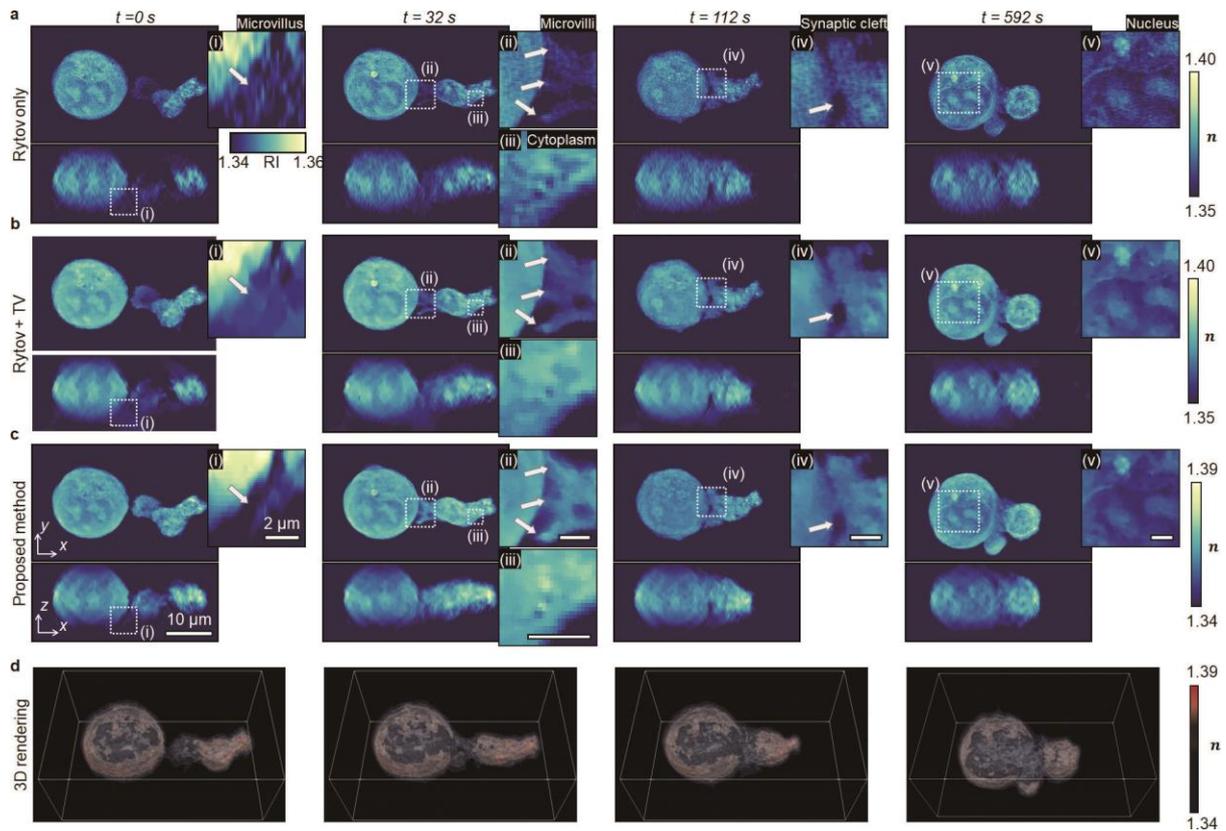

**Figure 3 | Reconstruction results of dynamic 3D cell–cell interactions between chimeric antigen receptor T and K562-CD19 cells.**

**a–c,** Cross-sections of tomograms reconstructed by (**a**) the Rytov approximation. (**b**) Rytov approximation with TV regularisation, and (**c**) our proposed method. Inset images: magnified images of white dashed squares. **d**, 3D rendered images of the results in **c**. The white arrows indicate either microvilli or a synaptic cleft.

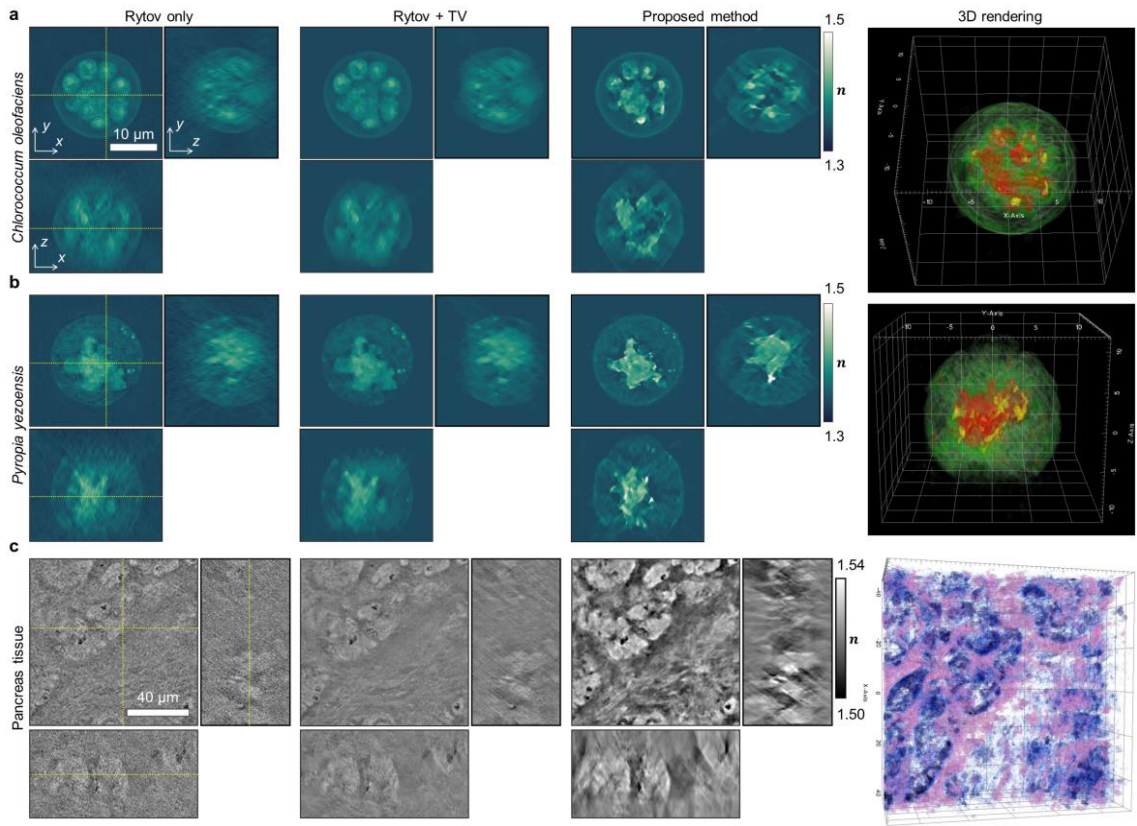

**Figure 4 | Reconstruction results of optically thick specimens**

**a–c,** Reconstruction results of (**a**) *Chlorococcum oleofaciens*, (**b**) *Pyropia yezoensis*, and (**c**) a 200-μm-thick human pancreas tissue. Each column represents cross-sectional images reconstructed by the Rytov approximation (first column), Rytov approximation with TV regularization (second column), and our proposed method (third column). The last column represents 3D rendered images of the tomograms reconstructed by the proposed method.

| **Inversion algorithm** |
|---|
| **y** = measured fields; **v'** = Rytov RI estimate; $\Psi_{in}$ = illumination fields |
| **while** not converged<br>    [**y'**, $\Psi'_{3D}$] = simulate($\Psi_{in}$, **v'**)<br>    [~, $\xi_{err,\ 3D}$] = simulate(**y'** – **y**, flip(**v'***,3))<br>    gradient = i/(8$\pi^2$) $\Psi'_{3D}$* $\odot$ flip($\xi_{err\ 3D}$, 3)<br>    **v'** = optimize(**v'**, gradient) |

**Table 1 | Pseudocode.**

| Sample | Type | | Wavelength (nm) | Objective NA | Scan NA ($n_m\sin\theta_{inc,max}$) | Step size ($\alpha$) | TV parameter ($\tau$) | TV iterations | Elapse time per iteration (s) | Total iterations | Used GPU |
|---|---|---|---|---|---|---|---|---|---|---|---|
| Single bead | Simulation 161 ×161 × 61 | Rytov + TV | 532 | 1.2 | 1.14 | 0.01 | 0.05 | 400 | 1.09 | 200 | GTX 1080 Ti × 7 |
| | | Our method | | | | 0.01 | 0.0075 | | 18.3 | | |
| | Experiment 220 ×220 × 71 | Rytov + TV | | | 1.07 | 0.05 | 0.075 | | 1.24 | | |
| | | Our method | | | | 0.0025 | 0.075 | | 22.0 | | |
| Dimer | Simulation 201 ×201 × 101 | Rytov + TV | | | 1.14 | 0.01 | 0.05 | | 1.07 | | |
| | | Our method | | | | 0.01 | 0.025 | | 49.1 | | |
| | Experiment 260 ×260 × 121 | Rytov + TV | | | 1.07 | 0.01 | 0.075 | | 1.99 | | |
| | | Our method | | | | 0.01 | 0.1 | | 51.8 | | |
| Trimer | Simulation 241 ×241 × 111 | Rytov + TV | | | 1.14 | 0.01 | 0.025 | | 1.56 | | |
| | | Our method | | | | 0.0075 | 0.0075 | | 90.4 | | |
| | Experiment 260 ×260 × 121 | Rytov + TV | | | 1.07 | 0.01 | 0.075 | | 2.14 | | |
| | | Our method | | | | 0.0075 | 0.1 | | 46.5 | | |
| Pentamer | Simulation 281 ×281 × 135 | Rytov + TV | | | 1.14 | 0.001 | 0.01 | | 2.39 | | |
| | | Our method | | | | 0.0075 | 0.075 | | 139.4 | | |
| | Experiment 322 ×322 × 161 | Rytov + TV | | | 1.07 | 0.01 | 0.1 | | 3.77 | | |
| | | Our method | | | | 0.01 | 0.1 | | 125.3 | | |
| Red blood cell | Simulation 221 ×221 × 81 | Rytov + TV | | 1.4 | 1.14 | 0.0075 | 0.005 | | 0.995 | | |
| | | Our method | | | | 0.01 | 0.0025 | | 40.2 | | |
| | Experiment 250 ×250 × 101 | Rytov + TV | | | 1.07 | 0.01 | 0.075 | | 1.54 | | |
| | | Our method | | | | 0.0025 | 0.075 | | 41.42 | | |
| T cell dynamics | Experiment 362 ×362 × 161 | Rytov + TV | | 1.2 | 0.84 | 0.01 | 0.002 | 100 | 5.76 | 50 | GTX 1080 Ti × 4 |
| | | Our method | | | | 0.001 | 0.0075 | | 39.36 | | |
| Chlorococcum oleofaciens | Experiment 275 ×275 × 211 | Rytov + TV | | | | 0.01 | 0.005 | | 5.21 | 150 | GTX 1080 Ti × 7 |
| | | Our method | | | | 0.005 | 0.005 | | 89.54 | | |
| Pyropia yezoensis | Experiment 317 ×317 × 211 | Rytov + TV | | | | 0.01 | 0.005 | | 6.21 | | |
| | | Our method | | | | 0.005 | 0.0025 | | 145.71 | | |
| Human tissue | Experiment 154 ×154 × 501 | Rytov + TV | 457 | 1.1 | 0.98 | 0.001 | 0.0025 | 100 | 2.77 | 100 | GTX 1080 Ti × 4 |
| | | Our method | | | | 0.01 | 0.0005 | | 58.71 | | |

**Table 2 | Setup specifications and used parameters.**